\documentstyle[epsfig]{mn}
\newif\ifAMStwofonts
\ifoldfss
  \ifCUPmtlplainloaded \else
    \NewTextAlphabet{textbfit} {cmbxti10} {}
    \NewTextAlphabet{textbfss} {cmssbx10} {}
    \NewMathAlphabet{mathbfit} {cmbxti10} {} % for math mode
    \NewMathAlphabet{mathbfss} {cmssbx10} {} %  "   "    "
  \fi
  \ifAMStwofonts
    \ifCUPmtlplainloaded \else
      \NewSymbolFont{upmath} {eurm10}
      \NewSymbolFont{AMSa} {msam10}
      \NewMathSymbol{\upi}     {0}{upmath}{19}
      \NewMathSymbol{\umu}     {0}{upmath}{16}
      \NewMathSymbol{\upartial}{0}{upmath}{40}
      \NewMathSymbol{\leqslant}{3}{AMSa}{36}
      \NewMathSymbol{\geqslant}{3}{AMSa}{3E}

       \let\ge=\geqslant
    \fi
  \fi
\fi % End of OFSS

\ifnfssone
  \newmathalphabet{\mathit}
  \addtoversion{normal}{\mathit}{cmr}{m}{it}
  \addtoversion{bold}{\mathit}{cmr}{bx}{it}
  \newmathalphabet{\mathbfit} % math mode version of \textbfit{..}
  \addtoversion{normal}{\mathbfit}{cmr}{bx}{it}
  \addtoversion{bold}{\mathbfit}{cmr}{bx}{it}
  \newmathalphabet{\mathbfss} % math mode version of \textbfss{..}
  \addtoversion{normal}{\mathbfss}{cmss}{bx}{n}
  \addtoversion{bold}{\mathbfss}{cmss}{bx}{n}
  \ifAMStwofonts
    \ifCUPmtlplainloaded \else
      \UseAMStwoboldmath
      \makeatletter
      \new@mathgroup\upmath@group
      \define@mathgroup\mv@normal\upmath@group{eur}{m}{n}
      \define@mathgroup\mv@bold\upmath@group{eur}{b}{n}
      \edef\UPM{\hexnumber\upmath@group}
      \new@mathgroup\amsa@group
      \define@mathgroup\mv@normal\amsa@group{msa}{m}{n}
      \define@mathgroup\mv@bold\amsa@group{msa}{m}{n}
      \edef\AMSa{\hexnumber\amsa@group}
      \makeatother
      \mathchardef\upi="0\UPM19
      \mathchardef\umu="0\UPM16
      \mathchardef\upartial="0\UPM40
      \mathchardef\leqslant="3\AMSa36
      \mathchardef\geqslant="3\AMSa3E

       \let\ge=\geqslant
    \fi
  \fi
\fi % End of NFSS release 1

\ifnfsstwo
  \DeclareMathAlphabet{\mathbfit}{OT1}{cmr}{bx}{it}
  \SetMathAlphabet\mathbfit{bold}{OT1}{cmr}{bx}{it}
  \DeclareMathAlphabet{\mathbfss}{OT1}{cmss}{bx}{n}
  \SetMathAlphabet\mathbfss{bold}{OT1}{cmss}{bx}{n}
  \ifAMStwofonts
    \ifCUPmtlplainloaded \else
      \DeclareSymbolFont{UPM}{U}{eur}{m}{n}
      \SetSymbolFont{UPM}{bold}{U}{eur}{b}{n}
      \DeclareSymbolFont{AMSa}{U}{msa}{m}{n}
      \DeclareMathSymbol{\upi}{0}{UPM}{"19}
      \DeclareMathSymbol{\umu}{0}{UPM}{"16}
      \DeclareMathSymbol{\upartial}{0}{UPM}{"40}
      \DeclareMathSymbol{\leqslant}{3}{AMSa}{"36}
      \DeclareMathSymbol{\geqslant}{3}{AMSa}{"3E}

       \let\ge=\geqslant
    \fi
  \fi
\fi % End of NFSS release 2

\ifCUPmtlplainloaded \else
  \ifAMStwofonts \else % If no AMS fonts
    \def\upi{\pi}
    \def\umu{\mu}
    \def\upartial{\partial}
  \fi
\fi

\title[Time Dependent Modeling of 1997 Mrk 501 Data]
  {Time Dependent Modeling of the Markarian 501 X-ray and TeV Gamma-Ray Data 
   Taken During March and April, 1997}
\author[Krawczynski, Coppi \& Aharonian]
  {H.~Krawczynski$^{1,2}$,
  P.S.~Coppi$^1$, F.~Aharonian$^3$ \\
  $^1$ Yale University, P.O. Box 208101, New Haven, CT 06520-8101, USA\\
  $^2$ Now at: Washington University in St. Louis, Phys.\ Dept.,
 	1 Brookings Drive, Campus Box 1105, St. Louis, MO 63130, USA\\
  $^3$ Max Planck Institut f\"ur Kernphysik, Postfach 103980, 
       D-69029 Heidelberg, Germany}
\date{Accepted --- --. Received 2002, January 21}
\pubyear{2002}

\def\uka { \raisebox{-0.5ex} {\mbox{$\stackrel{<}{\scriptstyle \sim}$}}}

\begin{document}
\label{firstpage}
\maketitle
\begin{abstract}
If the high-energy emission from TeV blazars is produced by the Synchrotron Self-Compton (SSC) 
mechanism, then simultaneous X-ray and gamma-ray observations of these objects are a 
powerful probe of the electron (and positron) populations responsible for this emission. 
Understanding the emitting particle distributions and their temporal evolution in turn allows us to
probe physical conditions in the inner blazar jet and test, for example, various 
acceleration scenarios. Furthermore, by constraining the SSC emission model parameters, 
such observations enable us to predict the intrinsic (unabsorbed) gamma-ray energy spectra of these
sources, a major uncertainty in current attempts to use gamma-ray observations
to constrain the intensity of the Diffuse Extragalactic Background Radiation (DEBRA)
at optical/infrared wavelengths. 
As a next step in testing the SSC model and as a demonstration of the
potential power of coordinated X-ray and gamma-ray observations, 
we model in detail the X-ray and gamma-ray light curves of the 
TeV blazar Mrk 501 during its April-May 1997 outburst with a  
time dependent SSC model.
Extensive, quasi-simultaneous X-ray and gamma-ray coverage exists for this period. 
We discuss and explore quantitatively several of the flare scenarios presented in the 
literature. We show that simple two-component models (with a soft, steady X-ray 
component plus a variable SSC component) involving substantial pre-acceleration of 
electrons to Lorentz factors on the order of $\gamma_{\rm min}\,=$ 10$^5$ 
describe the data train surprisingly well. All considered models imply an emission region that is
strongly out of equipartition and low radiative efficiencies 
(ratio between kinetic jet luminosity and comoving radiative luminosity) 
of 1 per-mill and less. Degeneracy in both, model variant and jet parameters, 
prevents us to use the time resolved SSC calculations to substantially 
tighten the constrains on the amount of extragalactic gamma-ray extinction by the 
DEBRA in the relevant 0.5-50 microns wavelength range, compared to earlier work.
\end{abstract}
\section{Introduction}
\subsection{EGRET Blazar Observations}
The EGRET detector on board the {\it Compton Gamma-Ray Observatory}
showed that many blazars are copious gamma-ray emitters \cite{Hart:99},
their power at gamma-ray energies being comparable to
(for low luminosity sources, 
i.e.\ BL Lac objects) or dominating by a wide margin 
(for high luminosity sources, i.e., FSRQs, Flat Spectrum Radio Quasars, 
and OVVs, Optically Violantly Variables) the power emitted at longer 
wavelengths.
The nonthermal radiation component probably originates from a population 
of relativistic particles embedded in the collimated outflow (jet) 
from a super-massive ($10^6$ up to several times $10^9 M_\odot$) black hole.
The nonthermal continuum emission is commonly explained 
with Synchrotron Compton \cite{Ulri:97,Siko:01b} models: 
embedded in a jet which approaches the observer with  
relativistic velocity, a population of high energy electrons emits Synchrotron 
radiation at longer wavelengths and at shorter wavelengths, Inverse Compton (IC) 
radiation of high energy electrons off lower energy seed photons.
The origin of the seed photons is still uncertain
(e.g.\ B{\l}a\v{z}ejowski et al.\ 2000). 
The seed photon source could be
``external'' to the jet, e.g., radiation scattered and reprocessed by ambient
matter in the Broad Line Region near the black hole, or infrared
radiation emitted by dust in the inner nucleus of the host galaxy (External Compton
models).
Alternatively, the dominant seed photons are synchrotron photons from the same 
electron population responsible for the IC scattering 
(SSC, Synchrotron Self Compton models).
In a generic source, both external and internal seed
photons could be important in producing the observed spectrum. 
In the following we use the term Synchrotron Compton models if we
do not want to specify the source of the seed photons.

Alternative models, so-called ``hadronic'' models,
invoke hadronic interactions of a highly relativistic 
outflow which sweeps up ambient matter \cite{Pohl:00}, 
interactions of high energy protons with gas clouds moving across 
the jet \cite{Dar:97}, or, interactions of extremely high 
energy protons with ambient photons \cite{Mann:98}, 
with the jet magnetic field \cite{Ahar:00a}, or with both \cite{Muec:02}.
If the reported fluxes of the diffuse infrared background between 60 and 100
micron (Lagache et al.\ 1999, Finkbeiner et al.\ 2000) correctly describe
the DEBRA intensity in the far-infrared band, the "reconstructed"  spectrum of Mrk 501, 
corrected for intergalactic absorption, may contain a sharp pile-up at and above $15$~TeV. 
The latter cannot be explained by conventional Synchrotron Compton models. 
It has been argued that the presence of such a pile-up can be
explained by bulk-motion comptonization (in the deep Klein-Nishina regime)
of the ambient radiation by an ultra-relativistic 
conical cold outflow with a bulk Lorentz factor of $^>_\sim 10^7$, 
while the remaining part of the spectrum could be explained 
by a conventional SSC model (Aharonian et al.\ 2002).  

All these models have some degree of success in 
explaining the overall Spectral Energy Distribution (SED) of
gamma-ray blazars. 
However, one can break much of the apparent 
degeneracy between these models by taking advantage of the rapid, 
large-scale time variability these sources exhibit. 
Different models, for example, produce emission 
at a given frequency using particles of different energies, interaction
cross-sections, and cooling times. 
The response of different models to changes in source conditions 
or the injection of fresh new particles is therefore different 
and in principle distinguishable -- provided that
one has sufficient time resolution to fully sample the flux variations
and sufficient frequency coverage to constrain the different emission
components that may be present. 
%

%
% EGRET SUMMARY
%
In view of this potential payoff, considerable effort has 
been dedicated to carrying out multi-wavelength observations on
powerful EGRET blazars like 3C 279 \cite{Wehr:98}. 
While the campaigns have lent considerable support to Synchrotron Compton models, 
the results of the campaigns were not as conclusive as one might have hoped.
The reasons for this are three-fold:
% 1
\begin{enumerate}
\item
These blazars turned out to be highly variable on timescales down 
to at least hours \cite{Matt:97,Wagn:97}.
Even for the brightest objects, the instrument available for the 
gamma-ray observations, EGRET, simply did not have enough collection 
area to track all the gamma-ray flux variations, let alone provide 
high quality energy spectra.
% 2
\item In typical models the electrons responsible for the GeV EGRET 
IC flux emit their synchrotron radiation at $\sim$UV 
energies. 
However, UV observations are difficult if not impossible
because of atmospheric and galactic absorption. 
Thus the simultaneous observations that were made, e.g., 
at gamma-ray and X-ray energies, tracked radiation from electrons 
with very different energies and different cooling times and thus potentially 
different time histories and perhaps even emission regions.
% 3
\item
The observations showed that the gamma-ray emission in several 
EGRET blazars is not consistent with the SSC model, the simplest version
of Synchrotron Compton models (see e.g.\ the comprehensive modeling of 3C~279 broadband data 
described by Hartman et al.\ 2001). 
The necessity to consider in External Compton  
models alternative seed photon fields 
substantially complicates the unambiguous interpretation of the data, 
especially since along our line of sight the beamed emission from the 
jet often dominates, making direct observation of these other photon 
fields difficult. 
\end{enumerate}
\subsection{Potential of TeV Blazar Observations}
The second class of gamma-ray emitting blazars that 
EGRET discovered, the low power BL Lac objects like Mrk 421, 
were initially passed over as targets for extensive multi-wavelength 
campaigns since they were too weak in the EGRET band. 
The arrival of ground-based gamma-ray detectors like 
Whipple, HEGRA, and CAT with detection areas on the order of 
$10^5$~m$^2$, however, now allows us to follow their gamma-ray 
fluxes on minute timescales \cite{Gaid:96} and to routinely obtain 
detailed spectral information on timescales down to one hour \cite{Ahar:99a}. 
Besides their better accessibility at gamma-ray energies, 
these low power objects have several other important advantages. 
BL Lacs and their likely FR-I radio galaxy parent population 
appear to have underluminous accretion disks, i.e., ``external'' 
photon fields may not be important as seeds for IC scattering \cite{Chia:99a}.
This together with the fact that their time-averaged SEDs have successfully 
been described with one-component SSC models, strongly suggest that 
SSC models which have much fewer free parameters than External Compton models
indeed apply. 
Also, perhaps because of the lower internal and external radiation 
fields and thus lower radiative losses \cite{Ghis:98}, the characteristic 
electron energies appear to be higher for the lower power objects, 
moving their synchrotron emission peak out of the UV, squarely into 
the X-ray range, where individual flares strongly dominate the 
overall luminosity and can readily be observed with broad-band X-ray 
satellites like RXTE and BeppoSAX.''''
In the SSC model, the IC peak then moves from GeV to 
$\sim$TeV energies. Thus, simultaneous X-ray and TeV gamma-ray observations 
follow the evolution of the electron population responsible 
for the bulk of the source luminosity, and the observations are 
well-matched in the sense that they track the emission from the 
{\it same} electrons, providing tight constraints on the electron 
distribution and its time evolution.  SSC models that
apply to these objects are therefore testable, especially
with the next generation of X-ray and gamma-ray detectors coming
on line in the next few years. 

Proving whether an SSC model works or not has a potentially large payoff. 
If the model does not work, then we must significantly revise our 
understanding of the physical conditions and processes in these objects.
If it does work, then we can use it for example to probe the acceleration
processes at work in the innermost region of the jet. 
We can also use it to constrain the amount of extragalactic 
gamma-ray extinction due to pair production processes on the 
diffuse optical/infrared background 
$\gamma_{\rm TeV}\,$ + $\gamma_{\rm IR,o}$ $\rightarrow$ $e^+$ $e^-$
\cite{Goul:65,Stec:92}, by comparing the predicted intrinsic  
TeV gamma-ray energy spectrum with the observed one
(Coppi \& Aharonian 1999, Krawczynski et al.\ 2000, called 
``Paper I'' in the following'').  
Based on these considerations, the brightest TeV blazars, 
Mrk~421 ($z=0.031$) and Mrk~501 ($z=0.034$), have been the
subject of increasingly intensive observing campaigns.
This has led to the discovery of pronounced TeV gamma-ray / X-ray flux
correlations for Mrk~421 (Buckley et al.\ 1996, Takahashi et al.\ 1996,
Maraschi et al.\ 1999, Takahashi et al.\ 2000) 
and Mrk~501 (Pian et al.\ 1998, Djannati-Atai et al.\ 1999, 
Paper I, Sambruna et al. 2000, see also Fig.~\ref{xtcorr} of this paper). 
\subsection{Goal of this Paper and Relation to Previous Work}
The goal of this paper is to extend the analysis of 
Paper I, which was a first
joint analysis of an unprecedented set of X-ray/TeV 
monitoring data taken during the 1997 flare of Mrk 501. 
Using RXTE (X-ray) and HEGRA (TeV) observations
that were simultaneous to within a few hours (i.e.\ less than the
$\sim$12 hour characteristic variability time scale of the 
source), we showed that the X-ray flux of the source, 
particularly above 10 keV, was strongly correlated with the 
TeV flux, in accord with Synchrotron Compton models. 
Moreover, we found that for that two month period the data
were consistent with the quadratic relation expected
between the X-ray and TeV fluxes in a simple SSC model, although
a linear relation between X-ray and TeV flux plus a constant
base X-ray flux level also described the data satisfactorily.
We then used an one-zone, steady state SSC model to fit the 
X-ray/TeV energy spectra for several days in order to
see if the model could explain the data and to make
a first guess at the SSC source parameters. 
We found that the data could be fairly
well-described by a reasonable sequence of SSC models.  
By assuming that we were indeed seeing SSC emission and 
by taking (at the time) extreme values of the SSC model parameters 
(e.g., jet Doppler factors $\sim$100), we then placed constraints 
on the maximum amount of extragalactic gamma-ray absorption
present in the observed spectrum (Paper I, Fig.\ 10).
We will give an updated very detailed discussion of the implications of SSC 
and External Compton models on the intensity of the DEBRA
in a companion paper (Coppi et al.\ 2001). 

In this paper, we attempt to quantify how well SSC model predictions 
match observations by taking the 1997 data set and fitting the full observed 
spectral sequence using a time dependent SSC 
code \cite{Copp:92}. The code accurately models the temporal evolution 
of the energy distribution function of a population of relativistic 
electrons due to acceleration processes and radiative and adiabatic 
energy losses. However, it is a one-zone code that assumes homogeneous
and isotropic particle and pitch angles distributions in the jet rest frame. 
Leaving aside plasma and magnetohydrodynamics issues,
such assumptions are clearly an oversimplification given the 
inhomogeneous structure of jets, especially as observed on VLBI 
radio scales where several distinct ``blobs'' (emission regions) 
may be active at any given time.
However, during 1997 the Mrk~501 emission was strongly dominated by 
individual flares during which the X-ray flux increased by up to 5 times
and the TeV gamma-ray fluxes by up to 30 times.
It is highly probable that these individual flares were produced by single
jet regions, rather than being the superposition of several,
causally not connected events.
In addition, the physical conditions in all the emission regions 
seemed to be very similar: the X-ray and TeV gamma-ray fluxes were well 
correlated during more than two months (see Paper~I, and this paper 
Fig.\ \ref{xtcorr}), and the TeV energy spectrum stayed remarkably 
stable during more than 6 months (Aharonian 1999a-c).
As a final justification of our approach, the analysis presented 
in Paper~I showed that the X-ray and TeV gamma-ray fluxes typically varied 
on time scales of $\sim$1~day with shortest flux rise and decay times 
on the order of half a day. 
Kataoka et al.\ (2001) and Tanihata et al.\ (2001) analyzed RXTE and ASCA data taken 
during the years 1997-2000 and find, in accord with our results, that Mrk~501 has a low 
duty cycle for flares on time scales of a few hours and shorter.
Thus, the sampling of the 
data with 2 X-ray observations and several TeV gamma-ray observations per day was 
probably sufficient for giving a rough picture of how X-ray and TeV gamma-ray fluxes 
evolved in time. 

Observational signatures of Synchrotron Compton models have been described by 
various authors (see e.g.\ the references given in Table \ref{models}).
In the following we show for the first time an attempt to fit a prolonged 
sequence of X-ray and TeV gamma-ray data with a time dependent SSC code.
This approach makes it possible to use the full information encoded in the 
correlated flux variability at different wavelengths. 
In contrast to parametric SSC fits (see e.g.\ Paper I, Tavecchio et al.\ 2001)
the method uses a self-consistently evolved electron population which
assures that the assumed electron energy spectrum is {\it physically realizable}
from an initial acceleration spectrum 
(see also the discussion by Mastichiadis \& Kirk 1997).
We think that the approach of using a time resolved analysis 
to break model degeneracies will become increasingly more powerful 
and important as the sensitivity and energy coverage of X-ray and 
gamma-ray instruments continue to improve.
Note that a thorough understanding of the SSC model is also a necessary 
prerequisite for the evaluation of External Compton models which always include a
SSC component.

The rest of this paper is structured as follows.
In Sect.\ \ref{data} we introduce the data set
and show an updated version of the X-Ray/TeV gamma-ray flux correlation.
In Sect.\ \ref{code} we describe the model calculations 
and in Sect.~\ref{res} the time dependent model fits.
Finally, we discuss the results in Sect.~\ref{disc}.
\section{The Data Set}
\label{data}
During 1997 the BL Lac object Mrk~501 went into a remarkable state of
continuous strong flaring activity and the source was intensively 
monitored at X-rays and TeV gamma-rays.
During April and May, 1997 the source was regularly observed with the 
RXTE X-ray satellite, with typically two pointed observations per day
of $\sim$20~min duration (Paper~I).
Each pointing resulted in a high accuracy measurement of the
3-25~keV X-ray flux and photon index with an accuracy which was only
limited by systematic effects.
The curvature of the X-ray spectrum could be assessed for a couple 
of pointings with relatively high X-ray fluxes and long integration times.
On three days (April 7, 13th, and 16) the source was also scrutinized
with the BeppoSAX X-ray telescopes, revealing the X-ray energy spectrum
of the source over the broad energy range from 0.1 keV to $\sim200$ keV
\cite{Pian:98}. 

\begin{figure}
\hspace*{-0.1cm} \epsfig{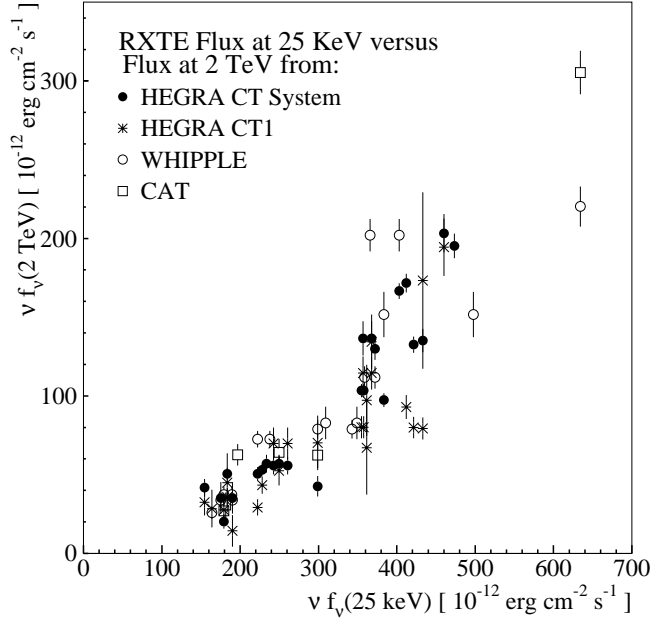}
\caption{\small 
Correlation between X-ray (RXTE) and TeV gamma-ray fluxes.
The gamma-ray fluxes are from CAT (squares), HEGRA CT~System (solid points), 
HEGRA~CT~1 (asterisks), and Whipple (open circles). Only observation 
pairs with less than 6~hrs time delay have been used.}
\label{xtcorr}
\end{figure}

In Paper~I we studied the correlation of the X-ray
fluxes with the TeV gamma-ray fluxes as measured with the HEGRA
Cherenkov telescope system \cite{Ahar:99a}. 
For the present study we complemented the data set with the
TeV fluxes from the HEGRA CT1 \cite{Ahar:99c}, Whipple \cite{Quin:99}, 
and CAT \cite{Djan:99} telescopes.
In Paper~I we found a very tight correlation between the
25~keV and 2~TeV fluxes. The flux variability amplitude 
was approximately 3 times larger at TeV than at X-ray energies, 
being consistent with a quadratic relationship.
An updated version of the X-ray/TeV gamma-ray flux correlation is
shown in Fig.\ \ref{xtcorr}. The additional X-ray/TeV gamma-ray flux 
pairs confirm the previous finding of a clear flux correlation. 
However, the quality of the correlation still does not allow us to 
differentiate between a quadratic X-ray/TeV gamma-ray relationship
and a linear one with a non-zero X-ray flux offset.

The X-ray as well as TeV gamma-ray data are plagued by systematic 
errors. In the case of the BeppoSAX data the spectral index below 1~keV
is not well determined due to uncertainties in the neutral hydrogen
column density. Above 50 keV the scatter of the data points increases
more than the statistical errors, indicating systematic uncertainties
in the detector response and/or the background subtraction procedure. 
Some TeV gamma-ray points taken at approximately the same time with 
different experiments deviate by more than 3~$\sigma$ statistical error 
from each other, indicative either for very fast source variability, or, 
for errors due to unstable atmospheric conditions. 
While the majority of nearly coincident measurements shows good
agreement between different TeV telescopes, the occurrence of
some exceptions makes it difficult to decide between models if
their predictions differ only for one or two days.
This caveat will be discussed further below, when we compare the
models with the data.

The spectral variability at TeV energies has been a matter of debate:
the HEGRA group did not detect spectral changes with an accuracy (1-5 TeV photon index)
of $\sim0.2$ and 0.05 for diurnal and flux selected mean energy spectra, respectively. 
The CAT group reported the statistically significant detection of a hardness intensity 
correlation based on the $F(>900{\rm \,GeV})/F(>450{\rm \,GeV})$ hardness ratio, 
corresponding to a $\simeq$0.25 change in photon index. 
The two data sets overlapped only partially in time:
the HEGRA group did i.e.\ not take data on April 16, 1997,
which is the most important day in the CAT analysis.
Konopelko et al.\ (1999) noted that the stability of the TeV energy spectra, 
evident in the HEGRA data, might be used to infer constraints on the 
intensity of the DEBRA.
In the plots shown below, we cross-calibrated the BeppoSAX data relative to 
RXTE measurements taken at approximately the same time.

Compared to the results shown by Pian et al.\ (1998)
we reduce the normalization of the BeppoSAX PDS data by up to 35\% which 
eliminates the discontinuity of the joint BeppoSAX MECS, LECS, and PDS 
energy spectra at $\simeq$15~keV (between the energy coverage of the 
LECS and PDS instruments) and is then consistent with the spectral shapes 
simultaneously measured from 3~keV to 25~keV with RXTE.
We also cross-calibrated the CAT, HEGRA CT1, and Whipple gamma-ray
fluxes relative to the ones measured by the HEGRA CT System.
Although we obtained a list of CAT fluxes as function of the integer 
MJDs of the observations, the fractional MJDs of 
the CAT observations are not known to us. 
In the following we centered the CAT observations at 12\,am UTC.
\section{Modeling}
\label{code}
\subsection{Time Dependent SSC Code}
The SSC code \cite{Copp:90,Copp:92} assumes a spherical emission 
region of radius $R$ which is filled with an isotropic electron population 
and a randomly oriented magnetic field $B$ and 
which approaches the observer relativistically.
The motion of the jet toward the observer can be characterized with 
the jet Doppler factor, defined by
\begin{equation}
\delta_{\rm j}^{-1}\,=\,\Gamma(1-\beta\,\cos{(\theta)}),
\end{equation}
with $\Gamma$ the bulk Lorentz factor of the emission plasma, 
and $\beta$ its bulk velocity in units of the speed of light, and
$\theta$ is the angle between jet axis and the line of sight in
the observer frame.
The TeV gamma-ray flux variability on time scale 
$\Delta T_{\rm obs}\,\approx$ 12~hr \cite{Ahar:99a}
together with causality arguments set an upper limit 
on the radius of the emission volume:
\begin{equation}
\label{r}
R\,\,\uka\,\,\delta_{\rm j}\,c\,\Delta T_{\rm obs}
\end{equation} 
If the jet moves along a curved path more rapid flares 
could result from a change of the Doppler factor as the jet's
radiation beam sweeps across the observer.
The kinetic equations, discretized in energy, take fully into account the 
non-continuous character of IC processes in the Klein-Nishina regime, 
and are evolved in time with a two step implicit scheme treating first the 
photon distribution and subsequently the electron distribution.
The length of time steps is chosen such that the number of photons and 
particles per energy bin changes per step by less than 20\%.
The kinetic equation for the photon density 
(per unit volume and energy) $n_{\gamma}$ reads:
\begin{equation}
\label{paolo2}
\frac{\partial n_{\gamma}}{\partial t}
\,=\,
q_{\gamma}\,-\,p_{\gamma}\,-\,
\frac{c}{R\,(1+\kappa)}\,n_\gamma
\end{equation}
where $q_\gamma\,d\epsilon$ and $p_\gamma\,d\epsilon$ 
are the rate of photons being produced into and out of the 
energy interval $\left[\epsilon,\epsilon+d\epsilon\right]$
due to electron-magnetic field, electron-photon 
and 2-photon interactions. 
The last term of the right hand side represents photons which
escape from the emission region. 
The factor $c\,R^{-1}$in the last term assures that the photon 
density approaches steady state values only with a rise/decay 
constant longer than the light crossing time. 
The factor $(1+\kappa(\gamma))$ parameterizes the modification of the
photon escape time by Compton processes \cite{Copp:92};
however, for all the models discussed in the following, 
we have always $\kappa \ll 1$.
The Klein-Nishina effect decisively influences the resulting
gamma-ray energy spectrum and proper modeling is imperative.

The kinetic equation of the electron (and possibly positron) 
density $n_{\rm e}$ reads:
\begin{equation}
\label{paolo1}
\frac{\partial n_{\rm e}}{\partial t}
\,=\,Q_{\rm e}\,\,
-\frac{\partial}{\partial \gamma}
\left[\dot{\gamma}_{\rm cont}\,\,n_{\rm e}\right]
\,+\,q_{\rm e}\,-\,p_{\rm e}\,
-\,\frac{n_{\rm e}}{t_{\rm e, esc}}
\end{equation}
with $Q_{\rm e}(\gamma,t)$ from Eq.~(\ref{eq1}), 
$\dot{\gamma}_{\rm cont}$ gives the decrease of an electron's 
Lorentz factor per unit time due to continuous energy losses, 
and $q_{\rm e}\,d\gamma$ and $p_{\rm e}\,d\gamma$ 
are the rate of particles being produced or 
scattered into and out of the Lorentz factor interval 
$\left[\gamma,\gamma+d\gamma\right]$ due to non-continuous energy 
loss processes, respectively. 
The last term of the right hand side represents an energy independent 
escape probability of electrons from the emission region.
To first order approximation our code takes the non-vanishing
source extension into account through the last term in Eq.\ \ref{paolo2}.
As a consequence, the code is able to describe flux variations even 
on time scales on the order of $R/c$ in a qualitatively correct way.
We limit ourselves in this paper to describe the time variable emission 
component with a one-zone SSC model.
A one-zone model is able to approximate multi-zone models
as long as the spatial gradients of the magnetic field and the non-linear 
components in the properly modified kinetic equations (\ref{paolo2}) and
(\ref{paolo1}) are small.
Our code can i.e.\ mimic ``linearized inhomogeneous models'' as discussed by 
Kirk, Rieger \& Mastichiadis (1998) and Chiaberge \& Ghisellini (1999).
While External Compton models can be dominantly linear, the SSC model is inherently 
non-linear: the synchrotron component directly follows the evolution of the electron 
population, but the IC component results from the interaction of the electron 
population with the self-produced synchrotron photons. 
Since electrons and synchrotron photons traverse the emission region
on a time scale of $R/c$, one expects that the IC component lags
the synchrotron component by approximately one light crossing time
\cite{Copp:99}. 
This is the most drastic time lag effect expected in the SSC model.
For Mrk 501 however no such time lag has been observed so far,
the upper limit being about 12~hrs \cite{Ahar:99a,Ahar:99c,Kraw:00,Samb:00}.
As long as instrumental resolutions do not permit to resolve this time lag,
we think it is safe to use only one component to describe the time variable 
emission.

We fit the full two months data train using a single emission volume.
As we will point out in the discussion, it might be that individual
flares (of durations on the order of $\sim$1~day) are produced by
independent emission regions. Upon flaring, a region would expand
adiabatically, and thus fade away quickly.
Even in this case, our model should give reasonable results for
two reasons: 
(i) as it turns out the best fitting models have particle escape times on 
the order of the flux variability time scale;
(ii) the tight X-ray/TeV gamma-ray flux correlation argues for
a very similar size of the emission regions.
As a consequence, each flare is produced by 
freshly accelerated electron populations and modeling the flares 
with one emission region gives similar results as using 
several disjunct emission regions.
\subsection{Treatment of Particle Acceleration}
Given the sparse observational sampling of our data set in time and wavelength, 
we did not embark on modeling the 
acceleration process in detail but used instead an ``external'' acceleration 
function. 
We parameterize the production rate of freshly accelerated particles 
as function of electron Lorentz factor $\gamma$, 
spectral index of particle acceleration $p$, 
normalization $Q_0(t)$, minimum Lorentz factor $\gamma_{\rm min}$,
and high energy cut-off $\gamma_{\rm max}(t)$ as follows:
\begin{equation}
\label{eq1}
Q_{\rm e}(\gamma,t)\,=\,\, 
Q_0(t)\,\gamma^{-p}\,\exp{(-\gamma/\gamma_{\rm max}(t))}\,
\Theta(\gamma-\gamma_{\rm min})
\end{equation} 
with $\Theta(x)\,=\,0$ for $x\,<\,0$ and $\Theta(x)\,=\,1$ for $x\,\ge\,0$.
We use the canonical value of $p\,=\,2$ expected for diffusive 
particle acceleration at strong shocks \cite{Bell:78,Blan:78} and
do not consider the ramifications arising from the non-linear modification 
of the shock structure due to the backreaction of accelerated particles 
\cite{Bell:87} and mildly or ultra-relativistic shock velocities 
\cite{Kirk:99a}. % ,Kirk:00,Acht:01}.

The low-energy cutoff in the spectrum of accelerated electrons
$\gamma_{\rm min}$ is a critical model parameter.
If the radiative cooling time of electrons with Lorentz factor 
$\gamma_{\rm min}$ is shorter than all the other characteristic
time scales of the system, the main break of the electron spectrum 
occurs at $\gamma_{\rm min}$. Thus, at high enough values ($\sim 10^5$),
$\gamma_{\rm min}$ determines the energies at which the synchrotron 
and IC SEDs peak.
On theoretical grounds one expects much lower values of between 1 and 
the proton to electron mass ratio $m_{\rm P}/m_{\rm e}\,=$ 1836
\cite{Hosh:92,Levi:96,McCl:97}. 
We will use in the following a relatively low value of $\gamma_{\rm min}\,=$ 1000
as the fiducial value and will discuss higher values at several points.

We characterize the acceleration luminosity $l_{\rm e}$ by the 
pair-compactness parameter \cite{Copp:92}:
\begin{equation}
l_{\rm e}\,\,=\,\,\frac{L_{\rm e}\,\sigma_{\mbox{\tiny T}}}{R\,m_{\rm e}\,c^3}
\,\,=\,\,\frac{8\pi\,R^2\,\sigma_{\mbox{\tiny T}}}{3\,c}\,\,
\int\,\gamma_{\rm e}\,Q(\gamma_{\rm e})\,d\gamma_{\rm e}
\label{compact}
\end{equation}
\subsection{Treatment of Extragalactic Extinction}
The TeV gamma-ray spectra are expected to be modified by extragalactic 
extinction due to pair production processes of the TeV gamma-rays with 
photons of the DEBRA. The uncertain DEBRA intensity in the relevant 0.5-50 microns 
wavelength range introduces a major uncertainty in the modeling of the source.
While earlier estimates of the DEBRA level predicted negligible extinction 
at gamma-ray energies below $\sim$1~TeV, more recent observational and 
theoretical efforts suggest that this might not be true \cite{Prim:01}.
We think that model estimates of the DEBRA still have not reached
the reliability that we should limit our computations to a specific
DEBRA model. Rather we will treat the modification of the TeV flux level and
energy spectrum as not fully constrained.
Clearly, the DEBRA extinction does not modify the relative TeV gamma-ray
flux variations and we use the information encoded in the 
relative flux changes by fitting the TeV gamma-ray fluxes subject to a 
common constant scaling factor $\xi$. 
At 2~TeV one expects a $\xi$-value of between 0.2 and 1. 
We have varied the $\xi$-values in this range, 
and the qualitative conclusions presented below are robust and do not depend 
on the exact value of $\xi$. For the detailed fits presented below, 
we will take $\xi\, \simeq$ 0.5, a value which seems to be preferred 
by recent observations and theoretical modeling. 

Due to rather large statistical errors on diurnal
gamma-ray photon indices we did not attempt to fit the variations of the
TeV energy spectra with the SSC code. 
We did check that the modeled TeV gamma-ray energy spectra are consistent 
with the observed one, taking into account that extragalactic extinction
only steepens the gamma-ray energy spectra. 
\subsection{Fitting Procedure}
The free parameters of our model are the radius of the emission volume $R$, 
the jet Doppler factor $\delta_{\rm j}$,
the mean magnetic field $B$, the escape time of relativistic electrons 
from the emission region $t_{\rm esc}$, the normalization of the electron 
acceleration rate $Q_0$, and the minimum and maximum Lorentz factors 
of accelerated particles $\gamma_{\rm min}$ and $\gamma_{\rm max}$.
We fit the April and May, 1997 RXTE 10~keV fluxes and 3-25~keV photon indices 
and the 2~TeV fluxes derived from CAT, HEGRA, and Whipple measurements.
Given a hypothesis of what causes the flaring activity (a variable 
$Q_0(t)$, $\gamma_{\rm max}(t)$, and/or $\delta_{\rm j}(t)$) 
we fit the data in 2 steps:
\begin{enumerate}
\item For a set of parameters 
($R$, $\bar{\delta_{\rm j}}$, $B$, $\gamma_{\rm min}$, 
$t_{\rm esc}$) we determine the simplest 
possible function $Q_0(t)$ and $\gamma_{\rm max}(t)$, 
or for some models $\delta_{\rm j}(t)$, which fit the X-ray flux amplitudes.
Hereby, ``the simplest possible functions'' means that given the
X-ray flux measurements at times $t_{\rm i}$ 
(in the jet frame) we use simple prescriptions to 
determine $Q_0$, $\gamma_{\rm max}$, or $\delta_{\rm j}$ at the times 
$t_{\rm i}'\,=$ $t_{\rm i}-\Delta t$, and compute intermediate values 
by simple interpolation.
The parameter $\Delta t$ is the time by which the photon density in the 
emission region reacts to changes of the electron spectrum.
In general, the optimal delay $\Delta t$ depends on the
time scale on which electrons cool and escape from the emission region.
Due to the observational constraint on the time lag between low (3~keV) and high energy 
X-rays (30~keV) to be shorter than $\sim$10~hrs \cite{Kraw:00}, 
the delay of all our models is dominated by the light crossing time 
and the treatment of Eq.~(\ref{paolo2}) results in an optimal value
of $\Delta t$ $\simeq\,2.5$ $R\,c^{-1}$.
For models in which only $\delta_{\rm j}(t)$ produces the time variability
we use $\Delta t\,=$ 0 for obvious reasons.

We determine the values $Q_0(t_{\rm i})$, $\gamma_{\rm max}(t_{\rm i})$, and
$\delta_{\rm j}(t_{\rm i})$ iteratively by making a first guess, 
computing the SSC model, and adjusting the values until the X-ray fluxes 
are described satisfactorily. Usually, between 2 and 5 iterations are needed. 
The reader should keep in mind that the true time history of the acceleration
process could be more complex.
\item We vary the parameters ($R$, $\bar{\delta_{\rm j}}$, $B$, $\delta_{\rm j}$, $t_{\rm esc})$
to obtain the best fit to the observed X-ray photon indices and TeV 
gamma-ray flux amplitudes.
The quality of the fits is characterized by $\chi^2$-values,
computed for the X-ray photon indices and for the TeV gamma-ray fluxes.
We exclude the first 2 days of each observation period from entering the
$\chi^2$-values, since the results strongly depend on the unknown behavior 
of the source before the observations commenced.
Note that the reduced $\chi^2$-values of the X-ray fluxes and photon indices 
and of the TeV gamma-ray fluxes exceed 1 by a wide margin, showing that
the experimental statistical errors are smaller than the accuracy 
of our models and/or that the experimental systematic errors
(often only poorly determined) are non-negligible.

Computing the $\chi^2$-values, we scale all TeV gamma-ray fluxes 
by a common factor $\xi$ with $0.2<\xi<1$.
For each model we state the $\xi$-value which we used as well as
$\eta$. The latter value is the difference of the observed and the modeled
mean spectral index. We interpret this difference as due to extragalactic
extinction. Due to the strong dependence of the IC luminosity on the radius of the
emission volume, the modeling does not give any constraints on $\xi$.
As will be discussed in Sect.~\ref{disc}, the uncertainty in the SSC model 
parameters do not allow us to constrain $\eta$ either.
\end{enumerate}
After obtaining in this way the best fits we test the
predicted SEDs (Spectral Energy Distributions) 
for consistency with the broadband X-ray spectra 
from the BeppoSAX observations, and an MeV upper limit 
from EGRET \cite{Cata:97}. 
\section{Results of the Time Dependent Modeling}
\label{res}
SSC blazar models have extensively been discussed in the literature.
The models can roughly be classified according to 2 criteria (see Table \ref{models}):
\begin{enumerate}
\item According to what produces the observed gamma-ray flares:
basically, almost every parameter of the SSC model has been invoked by at 
least one group to account for the blazar flaring activity.
\item According to the mechanism that determines the energies at which 
the synchrotron and IC SEDs peak. The SED peak energies are either 
determined by the minimum Lorentz factor $\gamma_{\rm min}$ of accelerated particles, or,
by the balance between radiative cooling times and the shorter of particle escape 
time and the characteristic duration of individual flares (sometimes referred to as
injection time scale, or, dynamical time scale of the jet).
\end{enumerate}
The modeling of the full data train is computationally very intensive 
and we therefore focused on exploring only the 
models which seemed most promising to us. 
While the time resolved analysis clearly rules out some models,
it gives fits of very similar quality for others.
Our difficulties to distinguish between models mainly derive from two facts:
\begin{enumerate}
\item From the limitations of the data set, namely sparse observational sampling
in time and energy, and systematic errors on X-ray energy spectra 
and TeV gamma-ray fluxes.
\item From the unknown modification of the TeV gamma-ray energy spectra
by extragalactic extinction.
\end{enumerate}
Keeping these limitations in mind, we discuss the fit results with a focus on 
pointing out which models are capable of correctly describing the qualitative 
behavior of the X-ray and TeV gamma-ray radiation.
Table \ref{parms2} lists the model parameters of the SSC models shown in the
figures.
\subsection{One-Component Models}
\subsubsection{Time variability through $Q_0(t)$}
\begin{figure}
\hspace*{-0.1cm} \epsfig{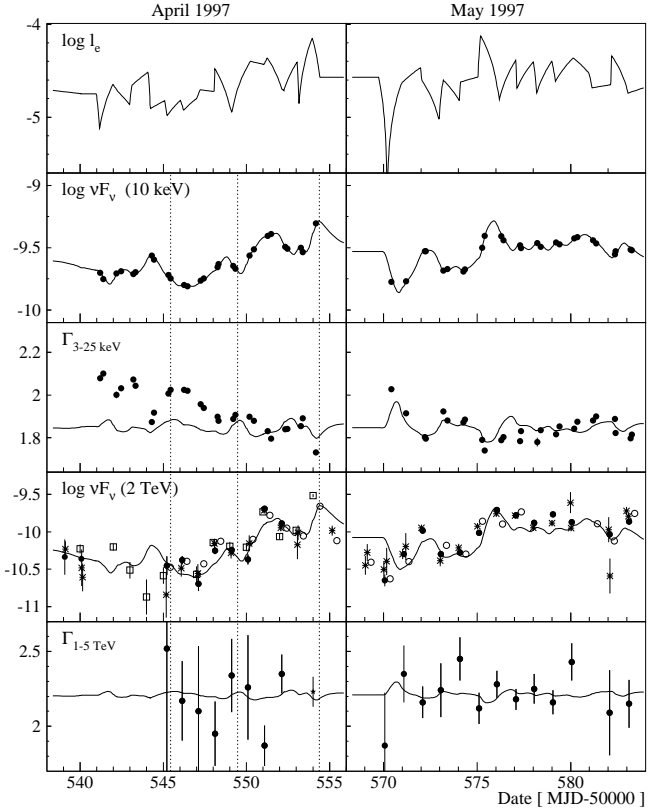}
\caption{\small 
X-ray and TeV gamma-ray data (data points) from April, 1997 (left column) 
and May, 1997 (right column) with SSC model fit (lines). 
Here, model flares are caused by a time dependent $Q_0(t)$.
The panels show from top to bottom: 
(i) the logarithm of the injection compactness $l_e$ 
(see Eq.\ \ref{compact}), 
(ii) the logarithm of the 10~keV X-ray energy flux 
(CGS units),
(iii) the X-ray 3-25~keV photon index,
(iv) the logarithm of the 2~TeV energy flux 
(CGS units), and
(v) the 1-5~TeV photon index.
The gamma-ray fluxes are from CAT (squares), 
HEGRA CT~System (solid points), HEGRA~CT~1 (asterisks), and 
Whipple (open circles). TeV photon indices have only been published by
the HEGRA group. The value for April 16 has been inferred from the
energy spectrum published by Djannati-Atai et al.\ (1999).
The vertical dotted lines show the days with BeppoSAX observations (April 7,
11, and 16 which will be discussed in more detail further below).
The model parameters are: 
$\delta_{\rm j}\,=$ 45, $R\,=$ 1.1$\times$ $10^{16}$~cm, 
$B\,=$ 0.014~G, $t_{\rm esc}\,=$ 10 $R\,c^{-1}$, 
$\gamma_{\rm min}\,=$ $10^3$,
$\gamma_{\rm max}\,=\,2.5 \times 10^7$, $\xi\,=$ 0.5, $\eta\,=$ 0.2.
}
\label{q0} 
\end{figure}

We first consider time variability through a varying rate of 
accelerated particles. 
If $Q_0(t)$ varies, the SSC mechanism automatically produces a 
more than linear increase of TeV flux as function of X-ray flux.
Fig.~\ref{q0} shows the observed and modeled X-ray and gamma-ray 
flux amplitudes and photon indices. 
In Paper I, we derived a lower limit on the Doppler factor of 6.3.
In most of the following models we use a rather high Doppler factor
of $\delta_{\rm j}\,=$ 45, for two reasons:
(i) a high Doppler factor allowed us to fit the data with a wide range 
of magnetic field values; for lower Doppler factors, weak magnetic fields 
result in a strong overproduction of TeV gamma-rays (see the related 
discussion by Krawczynski et al.\ 2001);
(ii) non-negligible extragalactic extinction of TeV gamma-rays seems highly probable;
the high Doppler factor results in predicted TeV energy spectra that agree
with the observed ones after correcting for extragalactic extinction
(which steepens the TeV gamma-ray energy spectra).
Ideally, we would like to use the modeling to determine the amount of extinction.
As we will discuss further below, for the time being we can not do this
due to parameter degeneracies. 

Whenever the size of the emission region satisfies Eq.~(\ref{r}), 
and either rapid particle losses ($t_{\rm esc}$ not much larger than 
$R\,c^{-1}$) or a sufficiently large magnetic field allow flares to 
decay rapidly enough, the X-ray amplitude can be described to 
arbitrary precision. This also applies for all the models described 
in the following. Thus, we subsequently focus on the X-ray photon indices and the 
TeV gamma-ray fluxes for measuring the quality of a fit.

Although the model describes the TeV flux levels, 
it fails to reproduce the range of observed X-ray photon indices.
The stability of the X-ray spectrum is a solid property of this
model. The spectral variability shown in Fig.\ \ref{q0} is already
the result of a fine tuning between the parameters $B$, $t_{\rm esc}$,
and $\gamma_{\rm max}$. Most realizations of this model
result in substantially less spectral variability.
For the model parameters of Fig.\ \ref{q0} the location of the break in
the synchrotron SED is given by the competition of the escape
of electrons on time scale $t_{\rm esc}$ and their radiative cooling through 
synchrotron and IC emission on time scale $t_{\rm rad}$.
High energy electrons with $t_{\rm rad}$ $^<_\sim$ $R$ $c^{-1}$ cool almost instantly.
Low energy electrons with $t_{\rm rad}$ $^>_\sim$ $t_{\rm esc}$ do not have 
time to cool before they escape the emission region. 
The result is that the spectrum does only vary over a rather small region 
where $t_{\rm rad}$ $\sim$ $t_{\rm esc}$. 
The ``smearing'' of the break in the electron spectrum due to the width of the 
synchrotron emissivity results in a rather stable break of the synchrotron spectrum.
Qualitatively, we see a similar behavior also for larger values of 
$t_{\rm esc}$. The reason is that the rise and decay times of the flares
are of the same order of magnitude as the time between flares. 
Thus, all observed electron energy spectra radiatively cool
to approximately the same degree with a break at approximately the same 
Lorentz factor.

Inspection of the fitted and observed TeV fluxes shows a strong 
discrepancy at MJD 50544. A TeV data point that suggests a very low 
flux is bracketed by X-ray observations of relatively high fluxes.
This model as well as the models described below fail to describe this 
exceptional anti-correlation. 
The low TeV flux may either be indicative of a short period of
low X-ray and TeV gamma-ray activity between the two X-ray observations, 
or of an underestimated TeV flux due to instrumental or atmospheric 
irregularities. Note how little the TeV photon index changes in this model.
\subsubsection{Time variability through \protect{$\gamma_{\rm max}(t)$}}
\begin{figure}
\hspace*{-0.1cm}\epsfig{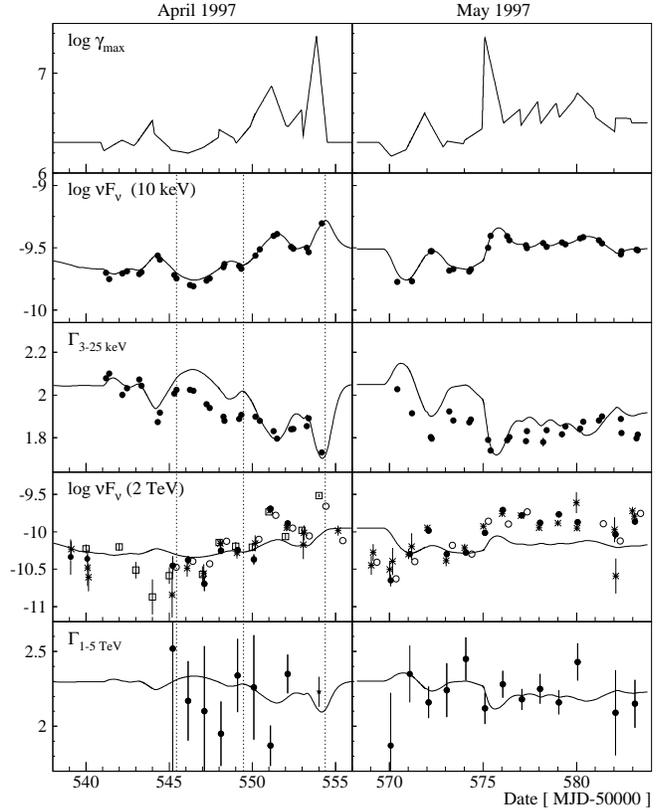}
\caption{\small 
Same as in Fig.~\ref{q0}, but with flaring activity through
a time dependent maximum Lorentz factor of accelerated particles
$\gamma_{\rm max}(t)$.
The upper panel shows here $\gamma_{\rm max}(t)$.
The model parameters are: 
$\delta_{\rm j}\,=$ 45, $R\,=$ 1.5$\times$ $10^{16}$~cm, 
$B\,=$ 0.009~G, $t_{\rm esc}\,=$ 3 $R\,c^{-1}$, 
$\gamma_{\rm min}\,=$ $10^3$,
$\xi\,=$ 0.5, $\eta\,=$ 0.2.
}
\label{gm} 
\end{figure}
As a second model we tested flares caused exclusively by a variation
of the high energy cutoff $\gamma_{\rm max}$ of accelerated particles.
While $\gamma_{\rm max}$ may depend on the details of the magnetic field 
structure in the surrounding of a particle accelerating shock, 
other parameters influencing mainly the acceleration of lower 
energy electrons could remain constant.
Historically, such models were motivated by observation of the
blazar Mrk~421 which showed dramatic X-ray and TeV gamma-ray flux variability 
accompanied by only minor optical flux variability. 

Varying $\gamma_{\rm max}$ alone we did not achieve a satisfactory 
fit to the data. A typical result is shown in Fig.\ \ref{gm}. 
As before the X-ray fluxes can be described to arbitrary precision.
While the model has no difficulty in producing the observed range of X-ray 
photon indices, it fails to describe the X-Ray / TeV gamma-ray flux
correlation: the predicted TeV gamma-ray fluxes hardly vary at all.
Combinations of Doppler factor and magnetic field where the TeV flux changes 
more strongly than the X-ray flux result in steeper than observed 
TeV energy spectra. The time variation of $\gamma_{\rm max}$ causes large 
flux and spectral variability only at energies $\gg$10 TeV where the inherent 
TeV energy spectrum is extremely soft (photon index $^>_\sim$ 2.5).
Extragalactic extinction can not remedy this shortcoming
since it is believed to steepen and not to soften the TeV energy spectra.

Tavecchio et al.\ (2001) studied parametric SSC model fits to  
Mrk~501 snapshot data and concluded that the maximum Lorentz 
factor of accelerated particles is mainly responsible for the flaring activity.
However, detailed inspection of their fit parameters shows that they
described the data by varying the break of the electron spectrum
rather than the high energy cutoff. Furthermore, the fits involve a
substantial variation of the normalization of the electron spectrum
(i.e.\ the acceleration rate).
\subsubsection{Time variability through $Q_0(t)$ and $\gamma_{\rm max}(t)$}
\label{gav}
\begin{figure}
\hspace*{-0.1cm}\epsfig{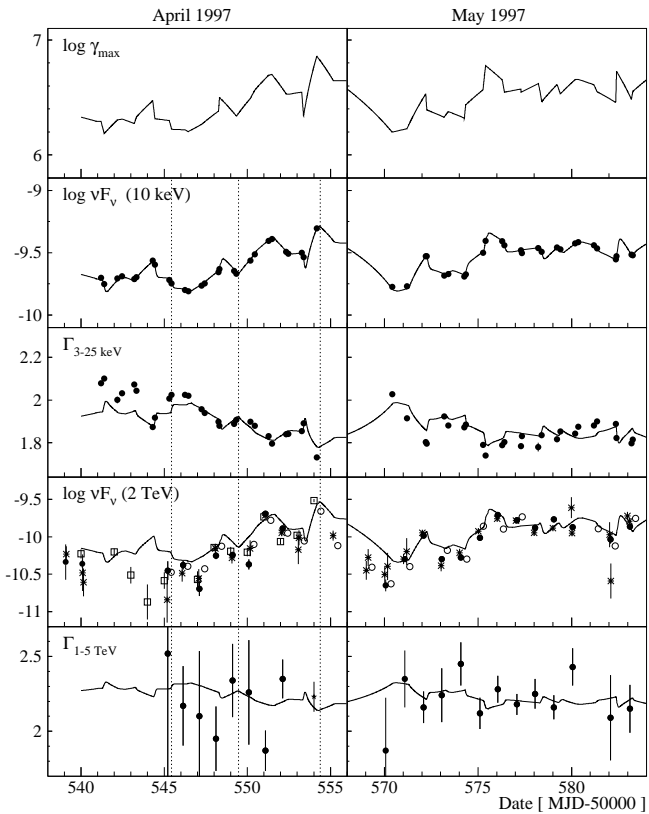}
\caption{\small 
Same as in Fig.~\ref{gm}, but with flaring activity through
time dependent $Q_0(t)$ and $\gamma_{\rm max}(t)$ with
\protect{$\gamma_{\rm max}\,\propto$} \protect{$Q_0\,^2$}. 
The upper panel shows here $Q_0(t)$.
The model parameters are: 
$\delta_{\rm j}\,=$ 45, $R\,=$ 3.2$\times$ $10^{15}$~cm, 
$B\,=$ 0.035~G, $t_{\rm esc}\,=$ 3 $R\,c^{-1}$, 
$\gamma_{\rm min}\,=$ $10^3$,
$\xi\,=$ 0.5, $\eta\,=$ 0.1.}
\label{alpha} 
\end{figure}

Models in which both, $Q_0$ and $\gamma_{\rm max}$, change with time 
have been invoked to account for the secular changes of the 
Mrk~501 X-ray \cite{Pian:98,Samb:00} and TeV gamma-ray \cite{Ahar:01a} 
energy spectra.
In models of diffusive electron acceleration at strong shocks the electron 
acceleration rate $Q_0$ is determined by the rate with which particles 
are ``injected'' into the acceleration process.
The high energy cutoff of accelerated electrons $\gamma_{\rm max}$
is determined by the competition between electron energy gains and 
energy losses. Changing plasma properties most probably affects both, 
the injection rate and the high energy cutoff.
Due to the uncertain nature of the particle injection mechanism we choose
a simple parametric description to describe the correlation between $Q_0$ and 
$\gamma_{\rm max}$:
\begin{equation}
\label{alphacorr}
\gamma_{\rm max}\,\,=\,\,Q_0\,^\alpha
\end{equation}
and treat the exponent $\alpha$ as an additional free parameter of the fit.
The $\chi^2$-value of the X-ray photon indices show a pronounced minimum
for a value of $\alpha\,=\,2$, and Fig.~\ref{alpha} shows a SSC fit to the data.
The model describes the RXTE and gamma-ray data rather satisfactorily.

During the first 3 days of the April campaign the model X-ray indices are
by $\sim$0.1 harder than the observed ones, a discrepancy which is shared
also by all subsequent models. Compared to other days of similar X-ray 
flux levels, the X-ray spectrum of the first three days with RXTE coverage 
was very soft, indicating that the source properties did evolve 
during the 2 months campaign.

\begin{figure}
\vspace*{0.1cm}
\begin{center}
\epsfig{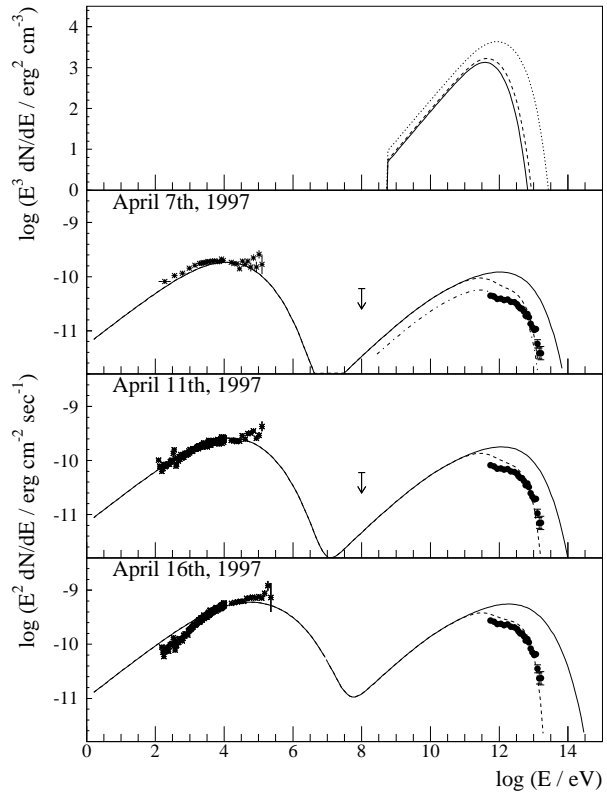}
\end{center}
\caption{\small 
For the model of Fig.\ \ref{alpha}
the upper panel shows the electron energy spectra 
($E^3\,dN/dE$, energy $E$ in the jet frame) 
responsible for the observations of April 7 (solid line), 
April 11 (dashed line) and April 16 (dotted line).
The lower three panels compare the observed (points) with the modeled 
(solid line) SEDs (energy in observer's frame).
For illustrative purposes the dashed line shows the TeV gamma-ray 
energy spectra modified by extragalactic absorption as predicted by 
the DEBRA model ``LCDM, Salpeter Stellar Initial Mass Function'' 
of Primack et al.\ (2001).
For April 7, the model under-predicts the TeV flux, and 
the dashed-dotted line shows the same absorbed gamma-ray spectrum, but normalized to the 
flux at 2 TeV to facilitate comparison of the spectral shapes.
Asterisks show the BeppoSAX data normalized to the flux measured 
with RXTE. The solid points show the shape of the HEGRA 1997 
time averaged Mrk~501 energy spectrum \protect\cite{Ahar:99b,Ahar:01b}
normalized to the mean TeV gamma-ray flux measured on each day
by CAT, HEGRA, and Whipple. 
The 2$\sigma$ upper limit at 100~MeV has been derived from
EGRET observations between April 9th and April 15th, 1997 
under the assumption of a constant emission level \protect\cite{Cata:97}.
}
\label{bs2} 
\end{figure}

Although only 3 BeppoSAX observations were performed during 1997,
the data is very constraining since it covers the broad energy range
from 0.1~keV to $\sim$200~keV.
The long BeppoSAX pointings of $\simeq$12~hrs duration
bracketed the RXTE and TeV gamma-ray observations. 
The comparison of the modeled and observed broadband energy spectra
is shown in Fig.\ \ref{bs2}. 
For all three days with BeppoSAX data the model fails to predict to 
observed X-ray energy spectra at and above 50~keV: the modeled
energy spectra are all too soft.
The high energy spectrum however is a rather solid prediction of this
model: the cutoff at relatively low energies is needed to correctly
account for the spectral variability observed with RXTE.

In Fig.\ \ref{bs2} also the modeled and observed TeV energy spectra are shown.
We used for all three days the time averaged 1997 HEGRA TeV gamma-ray 
energy spectrum normalized at 2~TeV to the mean flux measured on that day
with all operational TeV telescopes. 
The use of the time averaged energy spectrum is justified 
by the fact that all the HEGRA data taken during 1997 are 
consistent with the shape of the time averaged energy spectrum 
(see however Djannati-Atai et al.\ (1999)).
For illustrative purposes the dashed lines in Fig.\ \ref{bs2} show 
the predicted TeV gamma-ray  energy spectra modified by extragalactic 
absorption. 
We will further discuss the agreement between observed and modeled SEDs
below for the models which give a better overall fit to the X-ray data.
\subsection{Two-Component Models}
\label{2comp}
\begin{figure}
\hspace*{-0.1cm}\epsfig{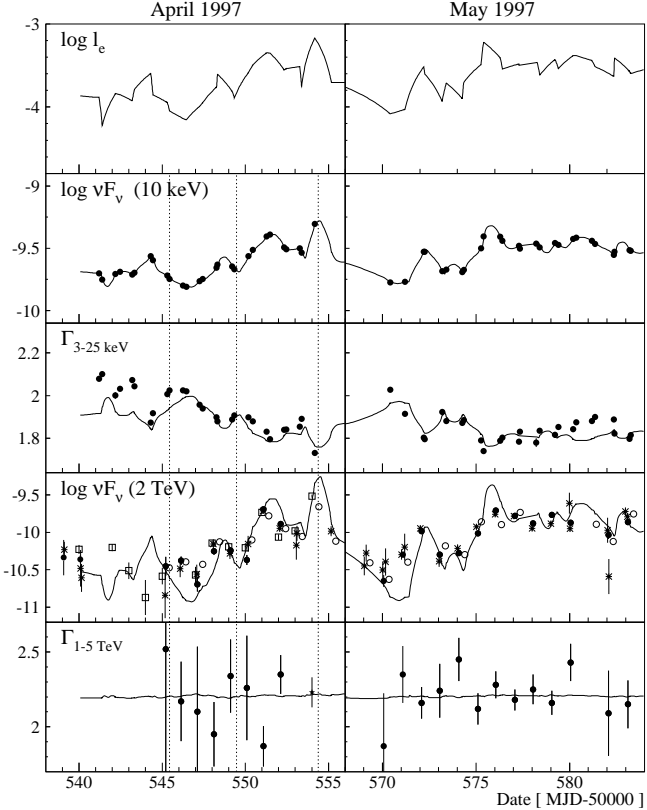}
\caption{\small 
Fit of a SSC model with two emission components:
(i) a quasi-stationary X-ray component, and, 
(ii) a time variable X-ray/TeV gamma-ray component, 
flares produced through $Q_0(t)$.
Data and units are the same as in Fig.\ \ref{q0}.
The model parameters are: 
$\delta_{\rm j}\,=$ 45, $R\,=$ 3.4$\times$ $10^{15}$~cm, 
$B\,=$ 0.014~G, $t_{\rm esc}\,=$ 3 $R\,c^{-1}$, 
$\gamma_{\rm min}\,=$ $10^3$,
$\gamma_{\rm max}\,=\,2.3 \times 10^7$, 
$\xi=0.5$, $\eta=0.4$.}
\label{2c} 
\end{figure}
\begin{figure}
\vspace*{0.1cm}
\begin{center}
\epsfig{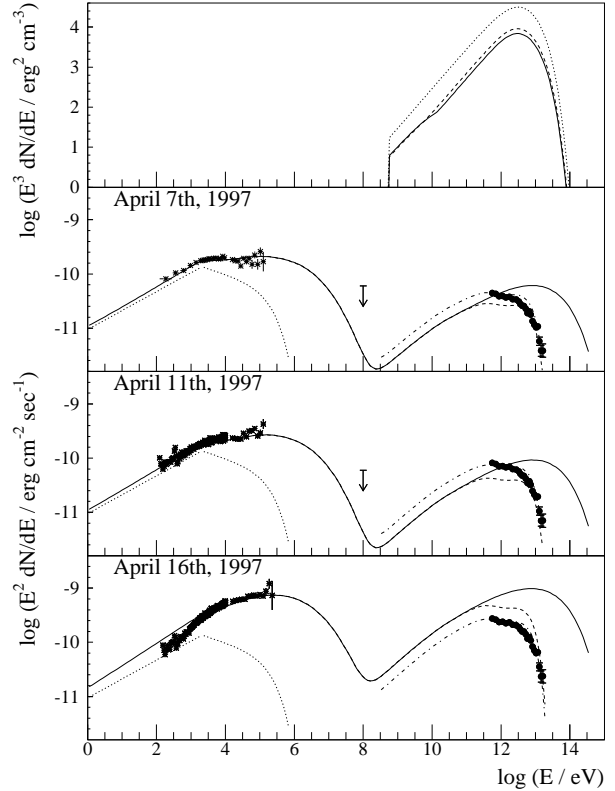}
\end{center}
\caption{\small 
Same as in Fig.\ \ref{bs2} but for the two-component model with time variable
$Q_0{\rm j}(t)$ shown in Fig.\ \ref{2c}. 
In the lower three panels, 
the dotted line shows the quasi-stationary X-ray component, 
and the dashed-dotted lines show the absorbed gamma-ray energy spectra
normalized at 2~TeV to the observed ones.
}
\label{bs1} 
\end{figure}
The region near the presumed central black hole as well as various 
regions along the jet might emit X-rays in Bremsstrahlung, IC, 
or Synchrotron processes without producing a comparable luminosity 
at gamma-rays (see also B{\l}a\v{z}ejowski et al.~(2000), 
Bicknell, Wagner \& Groves (2001)). Thus it is conceivable, even probable, 
that the X-ray emission from  Mrk~501 is ``contaminated'' 
by an emission component which varies on longer time scales than
the TeV gamma-ray radiation.
While all the one-component models described in the previous section 
failed to fully describe the data, 
we find that the addition of a quasi-stationary 
X-ray component substantially improves the situation.
Varying contributions of the quasi-stationary soft and the time variable
hard component are able to account for the large spectral changes observed
at X-rays.

Over the narrow spectral range form 3~keV to 25~keV we
describe the quasi-steady X-ray component by a power law.
We determined possible values of flux level and spectral 
slope of the quasi-stationary component from extrapolating the 
10~keV vs.\ 2~TeV flux correlation toward zero gamma-ray flux,
and the 10~keV flux vs.\ 3-25~keV photon index correlation (Paper I)
toward zero X-ray flux, respectively. 
Due to the scatter of both correlations this criterion gave a range of 
allowed values. We chose the values which resulted in the the best 
two-component SSC fits to the data, namely a 10 keV amplitude 
$\nu\,F_\nu\,=$ $10^{-10}$ erg cm$^{-2}$ s$^{-1}$ and a photon index of 2.2.

In the following we describe different incarnations of two-component models:
two with variability through a time dependent rate of accelerated particles,
and one with a time dependent Doppler factor of the SSC emission region.
\subsubsection{Time variability through $Q_0(t)$, $\gamma_{\rm min}\,=\,1000$}
\label{tcm}
First we consider a two-component model with flares caused by varying
$Q_0(t)$. Since the spectral variability at X-rays is produced by the
varying dominance of the soft quasi-static and the hard time dependent
components no additional spectral variability has to be produced by
a changing $\gamma_{\rm max}$ and we use a fixed $\gamma_{\rm max}$ 
corresponding to a high energy cut-off in the synchrotron spectrum 
at MeV energies.
Fig.~\ref{2c} shows the fit of the two-component model
(see figure caption for model parameters). 
Compared to the one-component $Q_0(t)$ model shown in Fig.\ \ref{q0}, 
the additional quasi-stationary soft component significantly 
improves the fit of the X-ray photon indices.
The improvement is most pronounced for the April data.
The model also describes well the TeV gamma-ray fluxes,
with the notorious exception of MJD 50544.

Finally, we compare the model SEDs with the one measured by BeppoSAX.
Based on the data of one of the three BeppoSAX observations 
(we used the observation of April 11), 
one can determine the spectrum of the quasi-stationary X-ray 
component outside the energy range covered by the RXTE observations. 
The other two observations can then be used to check the model predictions.
The upper panel of Fig.~\ref{bs1} shows the electron energy spectra
averaged over the integration time of the 3 BeppoSAX pointings.
The lower three panels compare the modeled with the observed SEDs. 
By construction, the model describes the X-ray data of April 7; 
the fit to April 11 is also good, but the model spectrum of April 16 
is too soft. More detailed inspection shows that the model fails to describe 
the temporal evolution of the $^<_\sim$ 1 keV fluxes, 
i.e.\ for April 16 it produces too much flux below 1~keV.

Remarkably, the predicted TeV energy spectrum, modified by 
extragalactic extinction according to the LSDM model of Primack et al.\ 2001
fits the HEGRA time averaged spectrum very well.
\subsubsection{Time variability through $Q_0(t)$, 
$\gamma_{\rm min}\,=$ $10^5$}
A similar model with a high value of $\gamma_{\rm min}\,\sim$ $10^5$ 
does not show these difficulties. 
In this case the break in the energy spectrum is more
abrupt and the peak of the synchrotron SED of the time variable component
is narrower than in the previous case.
Viable models with high minimum Lorentz factors are located in a 
completely different region of parameter space: at Doppler factor 45, 
we infer a magnetic field of $B\,=$ 1.1~G and a radius of 
$4.5\times$ $10^{13}$~cm compared to the values of $B\,=$ 0.014~G 
and $R\,=$ $3\times 10^{16}$~cm for the previous model.
A large magnetic field is required to assure sufficiently rapid cooling
of electrons with Lorentz factors above $\gamma_{\rm min}$ to
Lorentz factors below $\gamma_{\rm min}$. 
The latter electrons are needed to produce the optical and UV seed photons, and
partially, also for producing IC gamma-rays in the 250~GeV to $\sim 1$ TeV
energy range. A small radius $R$ follows than from the requirement to
produce the observed IC luminosity, given the large magnetic field and
the ``narrow'' synchrotron SED.
\begin{figure}
\hspace*{-0.1cm}\epsfig{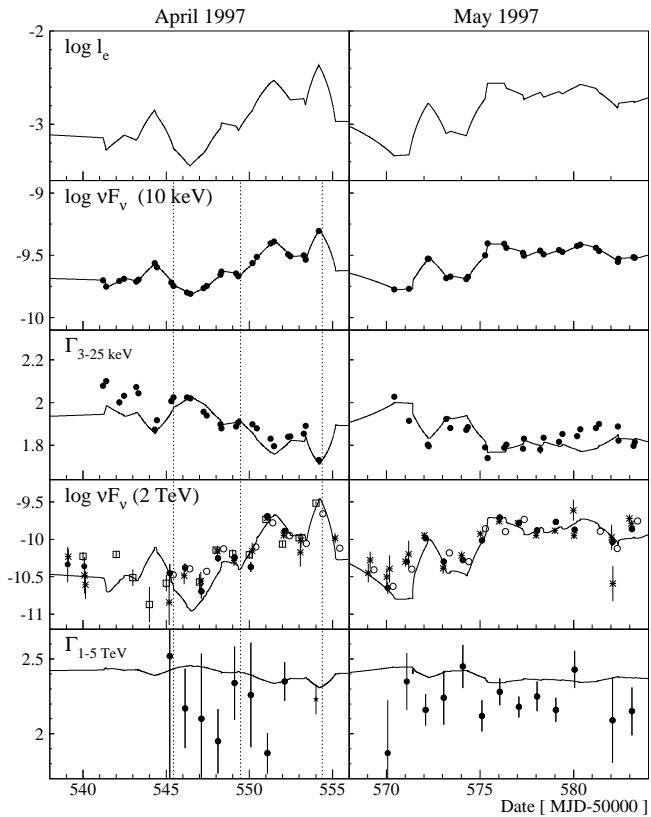}
\caption{\small 
Same as in Fig.\ \ref{2c}, but for the two-component model with 
$\gamma_{\rm min}\,=$ $10^5$.
The model parameters are: 
$\delta_{\rm j}\,=$ 45,
$R\,=$ 4.5$\times$ $10^{13}$~cm, 
$B\,=$ 1.1~G, $t_{\rm esc}\,=$ $10^4$ $R\,c^{-1}$, 
$\gamma_{\rm min}\,=$ $10^5$,
$\gamma_{\rm max}\,=$ $1.4\times 10^7$, 
$\xi =0.5$, $\eta =0.00$.}
\label{2cgm1} 
\end{figure}
\begin{figure}
\vspace*{0.1cm}
\begin{center}
\epsfig{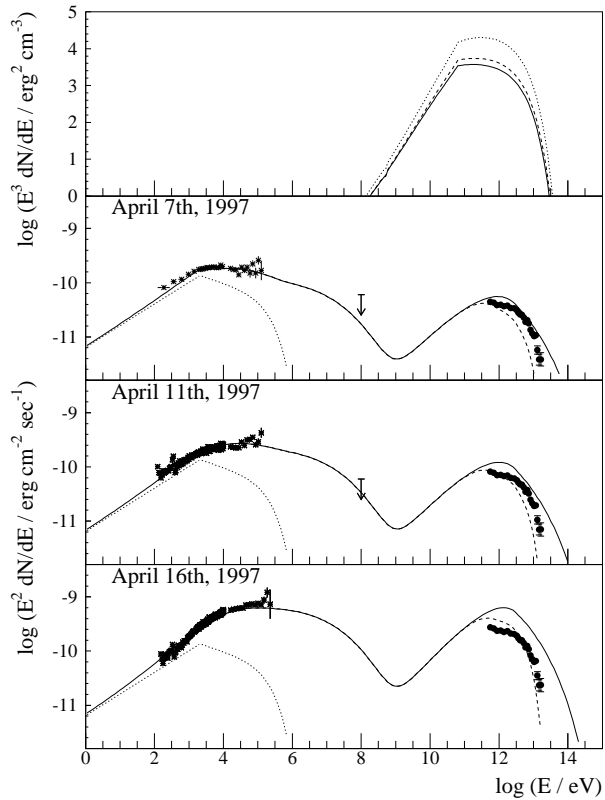}
\end{center}
\caption{\small 
Same as in Fig.\ \ref{bs1} but for the two-component model with 
$\gamma_{\rm min}\,=$ $10^5$, shown in Fig.\ \ref{2cgm1}. 
}
\label{2cgm2} 
\end{figure}

The best-fit result is shown in Figs.\ \ref{2cgm1} and \ref{2cgm2}.
In the latter, it can be recognized that the use of $\gamma_{\rm min}\,=$ $10^5$ 
substantially improves the fit to the broadband BeppoSAX data.
Even without any extragalactic extinction the model of the TeV gamma-ray
data is very soft and only barely consistent with the observed data 
below 10~TeV. Only above 10~TeV, the model implies a slight amount 
of extinction. Note the pronounced break of the IC spectrum at $\sim$2 TeV. 
Obviously, fitting a power law to a small portion of such a spectrum 
and inferring the DEBRA intensity from the deviation of the observed 
spectrum from this power law will not produce correct results.
\subsubsection{Variability through $\delta_{\rm j}(t)$, 
$\gamma_{\rm min}\,=$ $5\times 10^5$}
\label{dvar}
In the framework of one-component models, an emitting blob with constant and 
isotropic emission in its rest frame but with a varying angle between 
its motion and the line of sight can not account for the 1997 X-ray and 
TeV gamma-ray flares. 
The reason is that the large variability of the peak energy of the 
synchrotron SED would imply a large change of the blob's Doppler factor
and as a consequence a much larger than observed flux variability (Paper I).
In a two-component model however, a variable Doppler factor can explain 
the flux variability: the X-ray energy spectra mainly change due to the
relative dominance of the quasi-stationary and the time-variable 
X-ray components.
\begin{figure}
\hspace*{-0.1cm}\epsfig{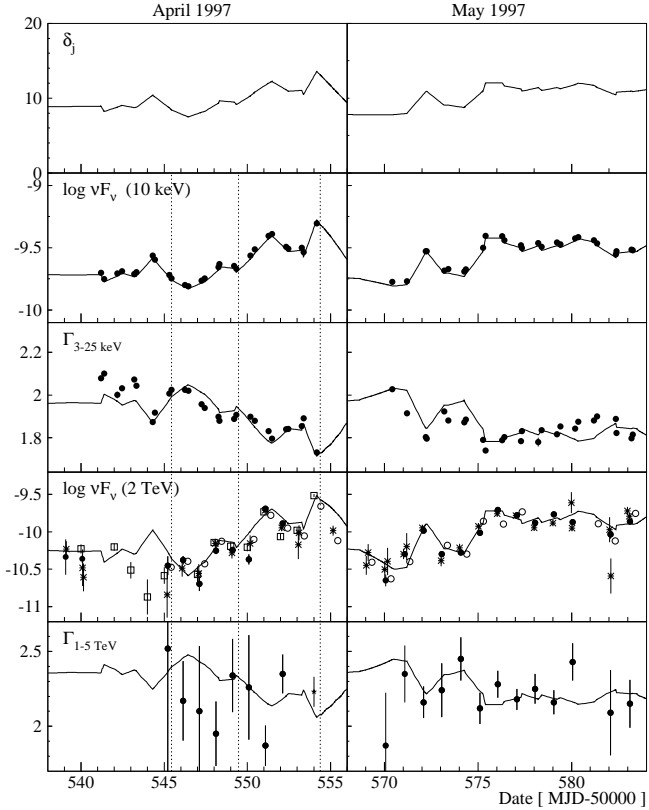}
\caption{\small 
Fit of a SSC model with two emission components:
(i) a quasi-stationary X-ray component, and, 
(ii) a time variable X-ray/TeV gamma-ray component, 
flares produced through $\delta_{\rm j}(t)$.
The upper panel shows here $\delta_{\rm j}(t)$, and the other
data and units are the same as in Fig.\ \ref{q0}.
The model parameters are: 
$R\,=$ $10^{15}$~cm, 
$B\,=$ 0.16~G, $t_{\rm esc}\,=$ 10 $R\,c^{-1}$, 
$\gamma_{\rm min}\,=$ $4.5 \times 10^5$,
$\gamma_{\rm max}\,=\,2.9 \times 10^7$, 
$\xi=0.5$, $\eta=0.04$.}
\label{dvar1} 
\end{figure}
\begin{figure}
\vspace*{0.1cm}
\begin{center}
\epsfig{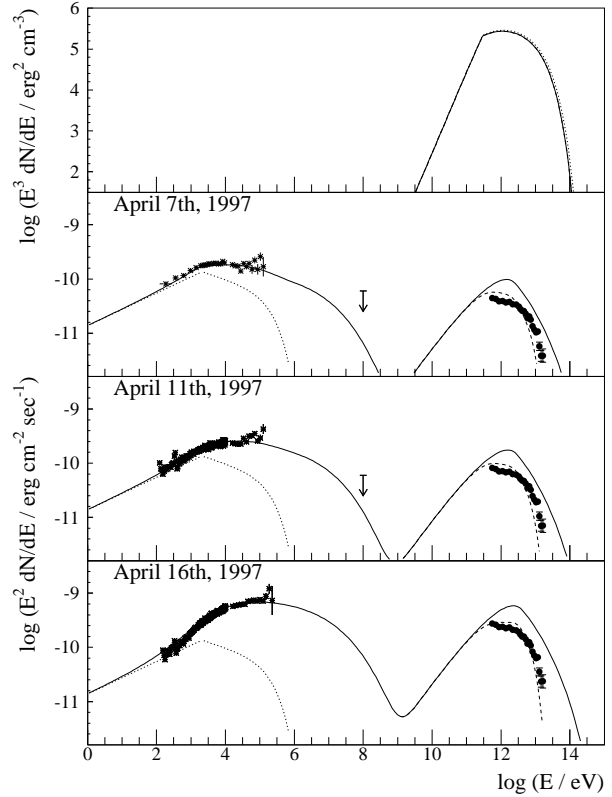}
\end{center}
\caption{\small 
Same as in Fig.\ \ref{bs1} but for the two-component model with time variable
$\delta_{\rm j}(t)$ shown in Fig.\ \ref{dvar1}. 
}
\label{dvar2} 
\end{figure}
Figs.\ \ref{dvar1} and \ref{dvar2} show the two-component fit to the 
time resolved data and the broadband spectral data, respectively.
The model gives an excellent fit to the data. 
This model is qualitatively very different from the other ones:
time variability can be produced on small time scales by changing $\delta_{\rm j}$, and
the electron spectrum is a steady state electron spectrum, and does not develop in time.
We used here a rather low value of the Doppler-factor, $\delta_{\rm j}\,=10$.
As a consequence, after correction for DEBRA extinction, the TeV energy spectra are 
steeper than the observed ones.
\section{Discussion}
\label{disc}
In this paper we describe the time resolved modeling of 
the X-Ray and TeV gamma-ray data of a 2 month observation campaign.
The time resolved analysis is plagued by the sparse observational sampling and
the unknown modification of the TeV gamma-ray energy spectrum 
by extragalactic extinction.
However, modeling the X-ray fluxes and energy spectra and
the relative changes of the TeV gamma-ray fluxes and photon indices
allows us to exclude some hypothesis about the flare origin.
Furthermore, we are able to verify that simple but self-consistently 
evolved SSC models based on canonical power-law energy spectra of 
accelerated electrons are able to account for the very detailed 
observational data.
More specifically, our conclusions from the time dependent modeling 
are as follows:
\begin{figure}
\hspace*{-0.1cm} \epsfig{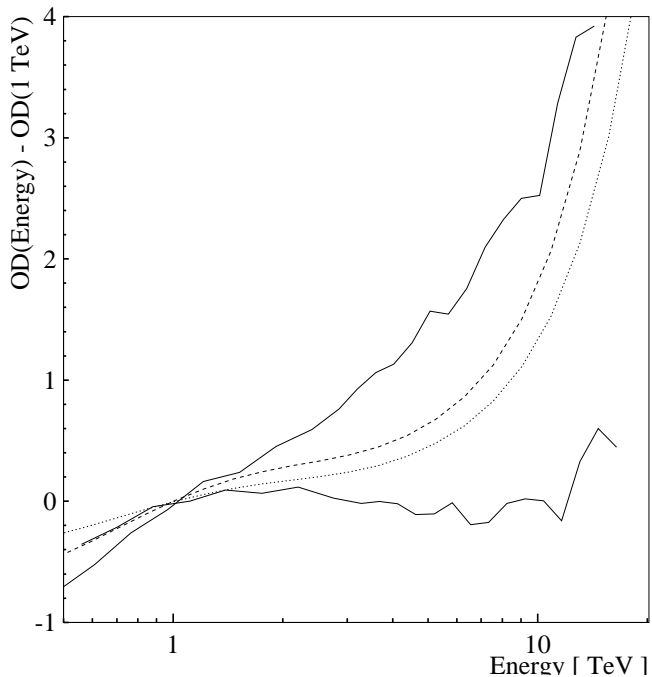}
\caption{\small 
Change of GeV/TeV gamma-ray optical depth as function of gamma-ray energy, 
inferred from comparing the model of Fig.\ \protect\ref{dvar2} 
(lower solid lines, respectively) with the 1997 time averaged 
gamma-ray energy spectrum measured by HEGRA (Aharonian et al.\ 1999b, 2001a).
The upper solid line is an upper limit for 
parameter combinations for which the 0.5-16~TeV energy spectrum 
approximates a power law of photon index 1.5. 
The upper limits include 2$\,\sigma$ statistical errors
and take the 15\% uncertainty of the absolute HEGRA energy 
calibration into account. 
Allowed DEBRA models have to lie between the two solid lines.
However, even if an absorption model lies between the two solid curves
it does not imply that a valid SSC model exists such that 
the absorbed gamma-ray spectrum describes the data
(the solid lines give a necessary but not sufficient condition
that a DEBRA model is consistent with the SSC models and the data).
The dashed and dotted lines show model predictions of Primack et al.\ 2001
(``LCDM'' with Kennicutt and Salpeter Stellar Initial Mass Functions, 
respectively).
All curves have been normalized to 1~TeV where the systematic errors on the
HEGRA energy spectrum are small. }
\label{debra} 
\end{figure}
\begin{enumerate}
\item One-component models do not fully describe the data.
While, by construction, the models succeed in accounting for the
temporal evolution of the X-ray fluxes they do not adequately
predict at the same time the range of observed X-ray spectral indices, the
broadband 0.1~keV-200~keV energy spectra, and the variation of the
TeV gamma-ray fluxes. 
\item Two-component models give surprisingly good fits to the data.
In these models, the X-rays originate from a superposition of a soft
quasi-steady component and a hard rapidly variable component.
We found two models which give an excellent fit: in the first model
flares are produced by a time dependent rate of accelerated particles.
In the second model, a changing Doppler factor causes the flares.
In both models, changes of the observed X-ray energy spectrum
mainly result from the relative dominance of the quasi-stationary
and the time-variable X-ray component. 
\item Accurate fits to the BeppoSAX broadband data 
require a large minimum Lorentz factor of accelerated 
particles on the order of $\gamma_{\rm min}\,=$ $10^5$. 
\item 
Degeneracy in both, model variant and jet parameters, 
prevents us to use the time resolved SSC calculations to substantially 
tighten the constraints on the amount of extragalactic TeV gamma-ray extinction, 
compared to earlier work (see e.g.\ Paper I, Guy et al.\ 2000, Vassiliev 2000,
de Jager \& Stecker 2002, and references therein).
The gamma-ray SEDs of Figs.\ \ref{bs1} and \ref{dvar2} 
are consistent with the LCDM DEBRA model of Primack et al.\ (2001).
In contrast, the model of Fig.\ \ref{2cgm2}, 
implies negligible extinction below $\sim$ 10~TeV. 
Especially the model with flux variability through the Doppler factor 
(Fig.\ \ref{dvar2}) can produce very different intrinsic gamma-ray SEDs 
while perfectly fitting the X-ray and gamma-ray flux variations and the
X-ray photon indices. 
In this model, the data constrain only the absolute flux level
and energy spectrum of the quasi-stationary component, and the relative
changes of the Doppler factor. The absolute value of $\delta_{\rm j}$,
as well as the parameters $R$, $B$, and $t_{\rm esc}$ remain degenerate.

An upper limit on the modification of the TeV gamma-ray energy spectrum
by extragalactic extinction can be derived from the fact, that the
emitted time averaged gamma-ray energy spectrum is unlikely to be 
harder than $dN_\gamma/dE\,\propto$ $E^{-\Gamma}$ with $\Gamma\approx 1.5$.
In Fig.\ \ref{debra} the range of allowed changes of gamma-ray optical 
depth with gamma-ray energy is shown and is compared to recent
model calculations of Primack et al.\ (2001).
A more accurate estimate of the amount of extragalactic extinction
from SSC modeling of Mrk~501 requires to pin down the jet parameters. 
Some key-observations are discussed further below; a more detailed 
discussion will be given by Coppi et al.\ (2002).
\item Table \ref{parms2} lists for all studied models
the electron to magnetic field energy density ratio
$r\,=$ $(u_{\rm e}\,/$ $u_{\rm B})$ as well as the
minimum kinetic luminosity $L_{\rm k}\,=$
$\pi$ $R^2$ $c$ $\Gamma^2$ $(u_{\rm e}\,+$ $u_{\rm B})$ 
\cite{Bege:94} and we use $\Gamma\,=$ $\delta_{\rm j}$.
All models are strongly out of equipartition with $r$ between 300 and 7500.
Similar results, derived from an one-zone stationary SSC model,
have recently been reported by Kino et al.\ (2002).
The kinetic luminosities lie between $5\times 10^{42}$ erg s$^{-1}$ 
and $2\times 10^{44}$ erg s$^{-1}$, about 1000 times and more than
the comoving radiative luminosities which are on the order of 
$\sim$ $5\times 10^{39}$ erg s$^{-1}$.
We computed the kinematic luminosity assuming a steady state jet
with the same physical parameters as the SSC emission region.
Although this assumption might overestimate the true kinematic
luminosity by a factor of a few, our models clearly indicate that
TeV blazars have rather powerful jets.
Models with high $\gamma_{\rm min}$-values are closest to 
equipartition and require the least power.
\end{enumerate}

In SSC models the X-ray to TeV gamma-ray luminosity ratio strongly 
depends on the size and magnetic field of the emission region.
As a consequence, most models that have been proposed to account 
for the flaring activity (as e.g.\ the internal shock model of 
Spada et al.\ 2001) do not naturally predict such a tight correlation
of X-ray and TeV gamma-ray fluxes through a large number of distinct 
flares as evident in Fig.\ \ref{xtcorr}.
The hypothesis that a single emission region of constant 
size produces a series of flares encounters several problems: 
(i) due to the strong dominance of particle pressure over magnetic field 
pressure, the emission region should quickly expand adiabatically and 
thus become undetectable; 
(ii) it is not clear how the energy required for sustaining
a prolonged flaring phase could be fed into the emission region;
(iii) during the flaring period that lasted more than 
$\Delta t\,\sim$ 2 months, the emission region would have advanced 
by $\sim c$ $\Gamma^2$ $\Delta t$, that means by a distance of 
about $\sim$100 pc. The stability of the radius of the emission region 
would thus imply a jet opening angle of $\sim 10^{-4}$ rad, 
several orders smaller than radio observations indicate.

Our preferred interpretation is that flares originate from distinct
emission regions with very similar characteristics, i.e.\ size and magnetic
field. Such emission regions might form as the jet becomes radiative at a
certain characteristic distance from the central engine.  
The fact that our models give particle escape times on the order of and 
shorter than the flux variability time scale indicates that the flare 
duration is limited by the adiabatic expansion of individual emission regions. 
The jet would naturally feed energy to the site where the flares originate.
The conclusions presented here are not limited to SSC models.
Also in External Compton models, the tight X-ray/TeV gamma-ray correlation indicates a 
preferred location for the production of individual flares: why else should 
the ratio of the jet frame magnetic field and external seed photon energy 
densities remain roughly constant during 2 months?

The preferred distance from the central engine could correspond
to a characteristic length at which the jet becomes unstable.
Alternatively, a change in ambient pressure could induce jet 
instabilities at a characteristic distance from the central engine. 
Note that a qualitatively different but similarly puzzling stability  
has been found in the hardness intensity correlation of 
Mrk 421 \cite{Foss:00} for measurements taken between days and years 
apart. One conclusion from this discussion is that refined
modeling should treat adiabatic expansion in more detail.

Since the modeling is computationally very intensive, we
explored only a limited number of models.
We did not consider External Compton models which historically have been 
applied to the more powerful EGRET blazars. 
For high jet Doppler factors even a very weak external photon field as 
e.g., IR radiation from dust, can be boosted and become significant 
in jet the frame. Depending on the seed photon energy spectrum, 
the radiative IC cooling of lower energy electrons might be stronger 
than for high energy electrons due to the Klein-Nishina effect. 
A possible consequence is that the energy spectra of External Compton models can be 
harder than for SSC models for a given value of $B$.
Thus, the Klein-Nishina effect introduces a very rich behavior
of External Compton models and the consequences of radiative cooling in 
the extremely ``blue`` TeV  gamma-ray blazars can substantially differ
from those in EGRET GeV blazars.

Crucial advances in fixing model parameters will only be possible
by substantially extending the observational coverage in 
time and wavelength.
The 1997 April and May observations had diurnal integration rates of 
typically 2 times 20 min. Pinning down the evolution of the source 
during several flares requires quasi-continuous monitoring over many days. 
Unfortunately, no sensitive X-ray all sky monitor with broadband
spectroscopic capabilities will be available for the next several 
years or even longer. Such an instrument would be able to 
participate in intensive Multiwavelength campaigns
on a large number of objects.
The upcoming generation of Cherenkov telescopes CANGAROO~III, H.E.S.S., 
MAGIC, and VERITAS will have energy thresholds of between 10~GeV and 50~GeV
and a one order of magnitude higher sensitivity than present day instruments.
The lower energy threshold is of crucial importance as it makes it
possible to asses the IC component at low energies where extragalactic
extinction is negligible ($z<0.1$) or much less ($z=0.5-1$) than at 
$\sim$ 500~GeV. The new experiments should be able to reliably assess 
changes of the diurnal GeV/TeV energy spectra with a statistical and 
systematic accuracy in photon index of 0.05 or better, due to better gamma-ray 
statistics and improved detector calibration and atmospheric monitoring.
Thus spectral changes as shown in Figs.\ \ref{gm}, \ref{alpha}, and
\ref{dvar1} will become measurable.

A key observation for fixing the jet parameters is to
measure a time lag between the X-ray and the TeV gamma-ray 
flux variability. A general prediction of SSC models is a time delay 
of approximately a light crossing time $R$ $c^{-1}$ between the 
leading X-ray and following gamma-ray fluxes.
The measurement of this delay would allow one to determine the
size of the emission region. The requirement that the DEBRA
reduces the 2~TeV flux by a factor of 5 or less would break the
degeneracy in $\delta_{\rm j}$ and $B$.
If the X-ray/TeV gamma-ray lag remains elusive it may be that the 
determination of the jet parameters of the TeV blazars detected so 
far has to wait until more reliable DEBRA estimates will be available
derived from multiple blazar detections at redshifts between 0.05 and 1.\\[2ex]
{\bf Acknowledgements}
The authors thank J. Quinn, D. Kranich, and A. Djannati-Atai
for the Whipple, HEGRA CT1, and CAT GeV/TeV data taken during 1997.
E. Pian kindly provided us with the BeppoSAX data.
We thank M.\~B\"ottcher, M.~Sikora, J.~Kirk, K.\ Motoki, and 
A.~Celotti for contributing very valuable comments.
HK thanks J.\ Primack for discussion on the
Diffuse Extragalactic Background Radiation
and for helpful suggestions regarding Fig.\ \ref{debra}.
An anonymous referee provided very good comments.
HK acknowledges support by the SAO (GO0-1169X and GO1-2135B).

\onecolumn
\begin{table}
 \caption{Selected Blazar SSC Models Relevant to This Paper}
 \label{models}
 \begin{tabular}{@{}p{4cm}p{2.5cm}p{1.3cm}p{3.5cm}lp{1.3cm}} \hline
 Authors & Objects Studied &  Time Dependent? & SED Peak Determined By & Flare Mechanism & DEBRA Extinction?  \\ \hline
Inoue \& Takahara (1996)      & 3C 279, Mrk 421 & No  & Cooling vs.\ Particle Escape & Not specified & No\\ 
Bednarek \& Protheroe 
(1997; 1999)                  & Mrk 421, Mrk 501& No  & Not specified & Not specified & Yes\\ 
B\"ottcher et al.\ (1997)     & Mrk 421         & No  & $\gamma_{\rm min}$ & $B$, $\gamma_{\rm min}$ & No\\ 
Mastichiadis \& Kirk (1997)   & Mrk 421         & Yes & Cooling vs.\ 
Particle Escape & $Q_0$, $\gamma_{\rm max}$, $B$ & No\\
Pian et al.\ (1997)           & Mrk 501         & No  & $\gamma_{\rm min}$ 
                                                      & $\gamma_{\rm min}$, $\gamma_{\rm max}$ & No\\
Dermer et al.\ (1998)         &generic          & Yes & Cooling vs.\ Particle Escape and 
                                                        Plasmon Deceleration& $\delta_{\rm j}$ & No\\
Chiaberge \& Ghisellini (1999)&generic          & Yes & Cooling vs.\ Particle Escape
                                                      & $Q_0$ & No\\
Coppi \& Aharonian (1999)     &generic          & Yes & Cooling vs.\ Particle Escape
                                                      & $Q_0$, $B$, $\gamma_{\rm max}$ & Yes\\
Kirk \& Mastichiadis (1999)   &generic          & Yes & Cooling vs.\ Particle Escape
                                                      & $Q_0$ & No\\ 
Kataoka et al.\ (2000)        & PKS 2155-304    & Yes & Cooling vs.\ Particle Escape
                                                      & $\gamma_{\rm max}$& No\\
Petry et al.\ (2000)          &Mrk 501          & No  & Cooling vs.\ Injection Time Scale
                                                      & $p$ & No\\
Kusunose et al.\ (2000)       &generic          & Yes & Cooling vs.\ Particle Escape
                                                      & $\gamma_{\rm max}$ 
                                                        (through $t_{\rm esc}$ and
                                                         $t_{\rm acc}$) & No\\
Tavecchio et al.\ (2001)      &Mrk 501          & No  & Not specified 
                                                      & Change of $\gamma_{\rm b}$ & No\\ 
Krawczynski et al. (2001)     & Mrk 421         &Yes  & Cooling vs.\ Particle Escape & $\gamma_{\rm max}$ & No\\

Sikora et al.\ (2001)         & 3C 279, PKS 1406-076
                                                &Yes & Cooling vs.\ Injection Time Scale 
                                                       or $\gamma_{\rm min}$
                                                     & $Q_0$ & No\\
Kino et al.\ (2002) & Mrk 421, Mrk 501, PKS 2155-304 
                    & No & Cooling vs.\ Particle Escape 
                    & Not specified & No\\
This work                     & Mrk 501
                                                &Yes & Cooling vs.\ Particle Escape 
                                                       or $\gamma_{\rm min}$
                                                     & $Q_0$, $\gamma_{\rm max}$, $\delta_{\rm j}$& Yes\\ \hline
\end{tabular}
\end{table}
\begin{table}
 \caption{Parameters of Models Shown in Figures}
 \label{parms2}
 \begin{tabular}{@{}p{2.5cm}p{2cm}llllllllll}
 Time Dependent & 
 Comments &
 $\bar{\delta_{\rm j}}$ & 
 $R$ $\left[\rm cm\right]$& 
 $B$ $\left[\,\mbox{G}\,\right]$ & 
 $t_{\rm esc}$&
 $\gamma_{\rm min}$ &
 $\gamma_{\rm max}$ & 
 $\xi$ &
 $\eta$ &
 $u_{\rm e}/u_{\rm B}$ & 
 $L_{\rm k}$\\ 
 Parameter& 
 &
 & 
 & 
 & 
 $\left[ R\,c^{-1}\right]$ &
  &
  & 
  &
  &
  & 
 $\left[\rm erg\,s^{-1}\right]$\\ 
\hline
%
% 1.06 => 8.9
$Q_0(t)$ & 1-component                 & 45 & 1.1$\times 10^{16}$~& 0.014 & 10 & 1000 & 2.5$\times 10^7$ & 0.5 & 0.2 & 660 & 1.2$\times10^{44}$\\ 
$\gamma_{\rm max}(t)$ & 1-component & 45 & 1.5$\times 10^{16}$~& 0.0089 & 3  & 1000 & 1.6-25$\times 10^6$ & 0.5 & 0.2 & 1200 & 1.7$\times10^{44}$\\ 
$\gamma_{\rm max}(t)\,\propto\,Q_0(t)^{\,2}$
& 1-component & 45 & 3.2$\times 10^{15}$~& 0.035 & 3  &1000 & 1.6-6.3$\times 10^6$ & 0.5 & 0.1 & 860 & 8.3$\times10^{43}$\\ 
$Q_0(t)$ & 2-component & 45 & 3.4$\times 10^{15}$~& 0.014 & 3& 1000 & 2.3$\times 10^7$ & 0.5 & 0.4 & 7470 & 1.3$\times10^{44}$\\ 
$Q_0(t)$ & 2-component, high $\gamma_{\rm min}$     & 45 & 4.5$\times 10^{13}$~& 1.12 & 10000 & 1.0$\times 10^5$  & 1.4$\times 10^7$ & 0.5 & 0.00 & 290 & 5.6$\times10^{42}$\\ 
$\delta_{\rm j}(t)$  & 2-component, high $\gamma_{\rm min}$ & 10 & 10$^{15}$  & 0.16 & 10 & 4.5$\times 10^5$ & 2.9 $\times 10^{7}$& 0.5 & 0.04 & 2970 & 2.8$\times10^{43}$\\ 

\hline
 \end{tabular}
 \medskip
%\hspace*{2cm} \\
% $^a$ For a given $R$, $Q_0$ is always chosen such as to produce the observed
%      mean X-ray flux; for models with time dependent $Q_0(t)$ see figures.\\ 
\end{table}
\twocolumn
\end{document}